\documentclass[5p,twocolumn]{elsarticle}

\usepackage{url}
\usepackage{verbatim}

\usepackage{amssymb}
\usepackage{multirow}
\usepackage[hyperindex,breaklinks]{hyperref} 

\usepackage{preprintcover}  
\PreprintCoverPaperTitle{Measurement of the high-mass Drell--Yan differential cross-section in {$pp$} collisions at {$\sqrt{s}=7$}~TeV
with the ATLAS detector}
\PreprintIdNumber{CERN-PH-EP-2012-064}  
\PreprintCoverAbstract{This Letter reports a measurement of the high-mass
Drell--Yan differential cross-section 
in proton-proton collisions 
at a centre-of-mass energy of 7~TeV at the LHC.
Based on an integrated luminosity of 
$4.9$~fb$^{-1}$, the differential cross-section in the $Z/\gamma^{\ast} \to e^+e^-$
channel is measured with the ATLAS detector as a function of the invariant mass, \mee, in the range $116 < \mee < 1500$ GeV, 
for a fiducial region in which both the electron and the positron have transverse momentum
$p_{\rm T} > 25$~GeV and pseudorapidity $|\eta|<2.5$.
A comparison is made to various event generators and to the predictions of perturbative
QCD calculations at next-to-next-to-leading order.}
\PreprintJournalName{PLB}  

\def\pt{\ensuremath{p_{\mathrm{T}}}} 

\newcommand{\mee}{\ensuremath{m_{ee}}}

\journal{Physics Letters B}

\begin{document}

\begin{frontmatter}



\title{Measurement of the high-mass Drell--Yan differential cross-section in {$pp$} collisions at {$\sqrt{s}=7$}~TeV
with the ATLAS detector}



\author{The ATLAS collaboration}

\address{}

\begin{abstract}
This Letter reports a measurement of the high-mass
Drell--Yan differential cross-section 
in proton-proton collisions 
at a centre-of-mass energy of 7~TeV at the LHC.
Based on an integrated luminosity of 
$4.9$~fb$^{-1}$, 
the differential cross-section in the $Z/\gamma^{\ast} \to e^+e^-$
channel 
is measured with the ATLAS detector as a function of the invariant mass, \mee, in the range $116 < \mee < 1500$ GeV, 
for a fiducial region 
in which both the electron and the positron have transverse momentum
$p_{\rm T} > 25$~GeV and pseudorapidity $|\eta|<2.5$.
A comparison is made to various event generators and to the predictions of perturbative
QCD calculations at next-to-next-to-leading order.
\end{abstract}


\end{frontmatter}



\section{Introduction}

At hadron colliders, the Drell--Yan (DY) process~\cite{bib:dyorig},
proceeding at tree level via the $s$-channel exchange of a virtual
photon or $Z$ boson, can produce $e^+ e^-$ pairs over a wide range of
invariant mass, \mee. 
The differential cross-section
${\rm{d}}\sigma / {\rm{d}} m_{ee} $  is described by perturbative QCD
(pQCD) calculations at next-to-next-to-leading order
(NNLO). 
Given the simple experimental signature and the low backgrounds, a 
small experimental uncertainty can be achieved on the measured \mee\ distribution
that allows for a precision test of pQCD.
The mass spectrum is also
sensitive to
the parton distribution functions (PDFs), in 
particular to the 
poorly known
distribution
of antiquarks 
at large $x$~\cite{mstw2008}, 
where $x$ can be interpreted, 
at leading order, as the fraction of the proton momentum carried by the interacting parton.
Additionally, the  
production of DY  $e^+ e^-$ pairs
is a source of background for other 
Standard Model (SM) measurements,
and the mass spectrum
may be modified by 
new physics phenomena.

The differential cross-section for DY  $e^+ e^-$ pair production in the high-mass range
has been reported previously by the CMS~\cite{bib:cmsdy}, CDF~\cite{bib:cdfdy} and 
D0~\cite{bib:d0dy} collaborations.
With the ATLAS detector, total and differential cross-sections in a mass window 
of $66 < \mee <116$~GeV have been measured using the 2010 dataset~\cite{bib:atlasz0}. 
In addition, searches for new physics in the high-mass range of 
the~\mee~distribution have been performed~\cite{bib:zprime,bib:zprimenew,bib:cinew} 
and no deviations from the SM expectation were observed. 
This Letter reports 
an extension of these previous analyses by providing 
a measurement of the DY cross-section,
fully corrected for detector effects,  as a function of the  $e^+ e^-$  invariant mass up to 1500~GeV. 
 The cross-section is reported in a phase space 
slightly extended with respect to
the fiducial acceptance of  the  $e^+$ and $e^-$.
The results are compared to NNLO pQCD calculations with next-to-leading-order (NLO) electroweak corrections
from the {\tt FEWZ 3.1}~\cite{fewz,fewz1} framework and to the predictions from three event generators.

\section{The ATLAS detector}

The ATLAS detector is described in detail in Ref.~\cite{DetectorPaper}. The two systems
most relevant to this analysis are the inner
tracking detector, surrounded by a superconducting solenoid providing a 2~T axial 
magnetic field, and the calorimeter. 
Charged-particle tracks and vertices are reconstructed with silicon pixel
and microstrip detectors covering the pseudorapidity\footnote{ATLAS uses a right-handed 
coordinate system with its origin at the nominal interaction point 
in the centre of the
detector and the $z$-axis coinciding with the axis of the beam pipe. 
The $x$-axis points from the interaction point to the centre of the LHC ring,
and the $y$-axis points upward. 
The pseudorapidity is defined in terms of the polar angle $\theta$ 
as $\eta = - \rm {ln}~\rm{tan}(\theta/2)$, and $\phi$ is the azimuthal 
angle around the beam pipe with respect to the $x$-axis.  The angular distance is defined 
as $\Delta R = \sqrt{(\Delta \eta)^2+(\Delta\phi)^2}$.
Transverse momentum and energy are defined as $p_{\rm {T}}= p \times \rm{sin}~\theta$ 
and $E_{\rm{T}}=E \times \rm{sin}~\theta$, respectively.} range $|\eta| < 2.5$ and a straw-tube transition-radiation tracker 
covering  $|\eta| < 2.0$. 
Within the region $|\eta| < 3.2$, electromagnetic calorimetry
is provided by barrel and endcap detectors
consisting of lead absorbers and liquid argon (LAr) as the active material,
with fine lateral and longitudinal segmentation within $|\eta| < 2.5$.
The hadronic calorimeter is based on steel/scintillator tiles in the central 
region ($|\eta| < 1.7$) 
while the hadronic endcap calorimeters ($1.5 < |\eta| < 3.2$) 
use copper/LAr.

A three-level trigger system is used to select events. The first level is implemented in
 custom electronics and is followed by two software-based trigger levels. In 2011 the total output rate
of events recorded for physics analysis was 200--300~Hz.

\section{Simulated samples}

Simulated data samples were generated in order to estimate backgrounds and correct the signal
for the detector resolution, efficiency and acceptance.
The  {\tt PYTHIA 6.426}~\cite{bib:pythia} and {\tt MC@NLO 4.02}~\cite{bib:mcatnlo} Monte Carlo (MC) 
generators were used to model the DY signal.  
In addition, {\tt SHERPA 1.3.1}~\cite{bib:sherpa} was used
 to produce signal samples with up to three additional partons, 
and the final result of the analysis is compared to the generator-level predictions from all three programs. 
{\tt MC@NLO} was also used to simulate the $t\bar{t}$ background,
while {\tt HERWIG 6.520}~\cite{bib:herwig} was used for the diboson ($WW$, $WZ$ or $ZZ$) backgrounds. 
{\tt MC@NLO} was interfaced to {\tt HERWIG} to model parton showers
and fragmentation processes, and to {\tt JIMMY 4.31}~\cite{bib:jimmy} for underlying event simulation.
All event generators were interfaced to {\tt PHOTOS 3.0}~\cite{photos}
to simulate QED final-state radiation (FSR), except for {\tt SHERPA} which 
uses the method of Ref.~\cite{Yennie:1961ad}. 

The {\tt PYTHIA} and {\tt HERWIG} samples were generated using the modified leading-order (LO**) PDF 
set {\tt MRSTMCal}~\cite{bib:mrstmcal1} following the recommendations 
of Ref.~\cite{bib:mrstmcal}, while the {\tt MC@NLO} samples used the NLO {\tt CT10}~\cite{bib:ct10} set.  The {\tt SHERPA} samples used the default
  {\tt CTEQ6L1}~\cite{bib:cteq6l1} PDF set of the generator.  

All MC events were generated at $\sqrt{s}=7$~TeV 
and include the full ATLAS 
detector simulation~\cite{athena} based on {\tt GEANT4}~\cite{geant4}.
Settings of MC parameters that describe properties of minimum bias events and the underlying event
were chosen based on results from previous ATLAS measurements~\cite{mc-tune}.
The effects 
of having on average nine 
interactions per bunch crossing (``pile-up'')
were accounted for by overlaying simulated minimum bias
events. To match the measured instantaneous luminosity
profile of the LHC, MC events were reweighted to yield the same
distribution of the mean number of interactions per bunch crossing as measured
in data.

Several corrections were applied to the simulated samples. 
The electron\footnote{In the following electron can mean either electron or positron.} energy resolution was corrected to match 
that observed in data,
following Ref.~\cite{bib:isem}. 
In addition, the efficiencies for electrons to pass 
requirements on the trigger, the reconstruction, and the particle identification in the MC
simulation were corrected by {\it scale factors}, defined as the ratio of the measured efficiency in data to that in the simulation.
 The {\tt PYTHIA} signal MC sample was reweighted at generator-level to a version that used an ATLAS tune
found to yield a good agreement with the transverse momentum distribution of the 
$Z$ boson observed in data~\cite{bib:zrw}. 
This procedure gives an adequate description of the
transverse momentum distribution for the high~\mee~region studied in this analysis.

The {\tt PYTHIA} and {\tt MC@NLO} signal predictions were reweighted to a NNLO pQCD calculation
with~\mee-dependent $K$-factors obtained from a modified version of {\tt PHOZPR}~\cite{bib:phozpr}.
Additionally, NLO electroweak corrections,  calculated using {\tt HORACE 3.1}~\cite{bib:horace}, were applied to the {\tt PYTHIA}  MC sample. 
The $t\bar{t}$ sample was rescaled to its inclusive near-NNLO cross-section prediction~\cite{bib:Moc08,bib:Ali11} 
and the diboson samples were normalised to NLO cross-sections calculated using {\tt MCFM}~\cite{Campbell:1999ah}.

\section{Event selection}

The analysis is based on the full 2011 data sample collected at $\sqrt{s} = 7$~TeV.
The data were selected online by a trigger that required two electromagnetic (EM) energy deposits each with
a transverse energy greater than 20~GeV.
Applying trigger and data-quality requirements
yields an integrated luminosity of $4.9\pm0.1$~fb$^{-1}$. Events from these $pp$ collisions 
are selected by requiring a collision vertex with at least three associated tracks, each 
with transverse momentum greater than 400~MeV. Events are then required to have 
at least two electron candidates as defined below.

Electron candidates are reconstructed from the energy deposits in the calorimeter matched 
to inner-detector tracks. 
An energy scale correction obtained from an {\it in situ} calibration, 
using $W/Z$ boson and $J/\psi$ meson decays, following the recipe of Ref.~\cite{bib:isem},
is applied to the data.
The electron candidates are required to have a transverse energy 
$E_{\rm T} > 25$~GeV and pseudorapidity $|\eta| < 2.47$, excluding the transition 
regions between the barrel and endcap calorimeters at
$1.37 < |\eta| < 1.52$. 
They must satisfy the ``medium'' identification  criteria based 
on shower shape and track-quality variables~\cite{bib:isem}
to provide rejection 
against jets, and have a hit in the innermost pixel layer to suppress background
from photon conversions. 

If an event contains more than two electron candidates passing the above selection, the two with highest $E_{\rm T}$
are chosen. To further reduce the background from jet production, the leading 
(highest $E_{\rm T}$) electron is required to be isolated by demanding that the sum 
of the transverse energy in the calorimeter cells in a cone of $\Delta R = 0.2$ around the electron direction 
is less than 7~GeV. This sum excludes the core of the electron energy deposition and is
corrected for the $E_{\rm T}$-dependent transverse shower leakage from the core, as well as for pile-up contributions.

After all selection requirements, a total of 
26\,844 
candidate events are found in the~\mee~range considered.
The dominant background contributions (6--16\% depending on \mee)
are from 
dijet and $W$+jets production, in which one or more
jets pass the electron selection criteria.  
The former includes multi-jet, heavy-flavour quark and $\gamma$ + jet production. 
The latter includes
pair-produced top quarks and single-top-quark production, 
where at least one electron comes from the misidentification of a jet or a heavy quark.
A data-driven method is used to evaluate the sum of these backgrounds.
The probability
of a jet to be misidentified as an electron 
 (the {\it fake rate}) is determined 
in an $E_{\rm {T}}$- and $\eta$-dependent way
from 
background-enriched samples
recorded by nine different inclusive jet triggers,
with $E_{\rm {T}}$ thresholds in the range 20--240~GeV.
In each of these jet-triggered samples, the fake rate is calculated as the fraction
of  electron candidates passing the ``loose'' identification requirement that also
pass the  ``medium'' requirement.
Events containing electron candidates from $W$ or $Z$ boson decays are first removed by dedicated cuts
in order to avoid bias from
real electron contamination: $W$ candidates are rejected by requiring low missing transverse energy and low 
transverse mass; and $Z$ candidates are rejected if they contain
two ``medium'' electrons.
A weighted average of the fake rates obtained from the nine jet samples
is then calculated. 
To estimate the total dijet plus $W$+jets background,
a factor derived from the averaged fake rate is applied to events that pass the signal selection 
but with one or both electron candidates
passing the ``loose'' identification requirement and
failing the ``medium'' requirement. 

\begin{figure}[t]
\begin{center}
\includegraphics[width=\columnwidth]{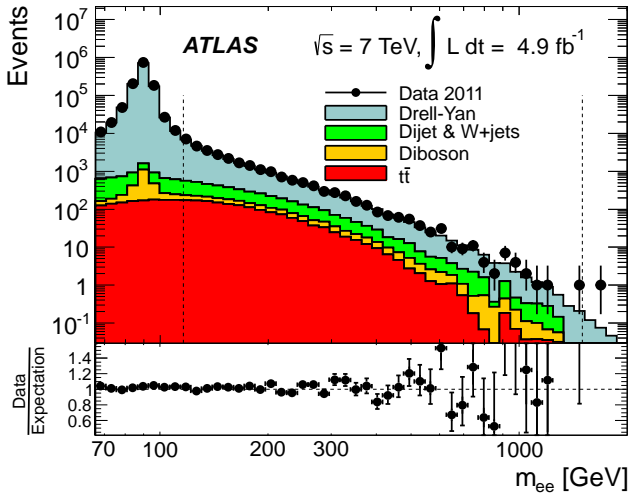}
\end{center}
\caption{Distribution of~\mee~in data compared to the summed signal and background predictions, where
the bin width is constant in log(\mee). The Drell-Yan signal is predicted from {\tt PYTHIA} simulation and the combined 
dijet and  $W$+jets contribution is estimated from data as described in the text. 
The dashed vertical lines indicate the mass range used for
 the differential cross-section measurement.}
\label{fig:Mee}
\end{figure}

The only other significant backgrounds arise from
the dileptonic decays of pair-produced top quarks ($t\bar{t}$) and from diboson production processes.
These contributions account for up to 5\% and 9\% of the selected events, respectively, and are
estimated from MC simulation.
The overall level of agreement between data and the sum of the MC signal and background predictions
(scaled to 4.9~fb$^{-1}$) and the data-driven 
background component is shown in Fig.~\ref{fig:Mee}.

\section{Cross-section measurement}

The differential cross-section, ${\rm{d}}\sigma / {\rm{d}}m_{ee}$, is measured in 13 bins of~\mee~from 116~GeV to 1500~GeV
in a fiducial region in which both electrons 
have transverse momentum $p_{\rm T} > 25$~GeV
and lie within $|\eta| < 2.5$. 
The cross-section and fiducial region are determined for two conventions
regarding QED FSR corrections. 
For the Born-level result, the true (meaning without detector simulation)~\mee~and electron kinematics are  
defined by the electrons originating from the $Z/\gamma^{\ast}$ decay before FSR.
At the {\it dressed} level, 
true final-state electrons after FSR are recombined with radiated photons within a cone of 
$\Delta R = 0.1$.

The cross-section is 
calculated from
\begin{equation} 
 \frac{{\rm{d}} \sigma}{{\rm{d}} \mee} = \frac{N_{\rm data} - N_{\rm bkg} }{C_{\rm {DY}}\,L_{\rm{int}}} \, \frac{1}{\Gamma_{\rm {bin}}},
\end{equation} 
where $N_{\rm data}$ is the number of candidate events observed in a given bin of~\mee\ (of width $\Gamma_{\rm {bin}}$), 
$N_{\rm bkg}$ is the 
total background in that bin and $L_{\rm{int}}$ is the integrated
luminosity.
The correction factor,  $C_{\rm {DY}}$, takes into account the efficiency of the signal selection and 
bin migration effects. 
It also includes the small extrapolation (about 10\% to 13\%)
over the small region in $|\eta|$ that is excluded for reconstructed 
electron candidates ($1.37<|\eta|<1.52$ and   $2.47<|\eta|<2.5$).
The correction factor is 
defined as the number
of MC-generated events that pass the signal selection in a bin of reconstructed~\mee, 
divided by
the total number of generated 
events within the fiducial region, at the Born
or dressed level, in the corresponding bin of true~\mee.
It is obtained from the {\tt PYTHIA} MC signal sample and corrected for differences in
the reconstruction, identification and trigger efficiencies between data
and MC simulation.
The value of $C_{\rm {DY}}$ varies from 0.55 (0.57) in the lowest bin to 
0.70 (0.73) in the highest bin at the Born (dressed) level.

The~\mee\ resolution varies from approximately 3\% at low \mee\ to 1\% at high \mee.
The purity,
defined as the fraction of simulated events 
reconstructed
in a 
given~\mee~bin that have true~\mee~in 
the same bin, ranges from 
79\% (82\%) to 98\% (98\%)
at the Born (dressed) level.

\section{Systematic uncertainties}
The main contributions to the 
systematic uncertainties are given in Table~\ref{tab:systSummary} and described below.

\paragraph{Background estimation}

In the estimation of the dominant dijet and $W$+jets background, a systematic uncertainty of 11\% is assigned to the 
$E_{\rm {T}}$- and $\eta$-dependent 
fake rate, corresponding to the spread of this quantity as measured in the nine independent jet samples,
in order to cover any possible bias introduced in the triggering of these background events.
A further uncertainty on the fake rate of up to 11\% arises due to the
presence of remaining 
signal contamination in the background-enriched sample.

The total systematic uncertainty on the fake rate combines 
with a smaller effect (around 5\%) from signal contamination in the sample
where the fake rate is applied, to give a total uncertainty on the 
resulting background estimate of up to 16\%.
An additional systematic uncertainty can arise if the fake rate differs
for different sources of fake electrons
and the relative contribution of the different sources is not the same
in the data sample where the fake rate is measured and the sample of events
to which it is applied.
It is found that $b$-jets have a
higher fake rate than
jets initiated by gluons or light quarks,
but that 
the fraction of $b$-jets is small and similar in both samples.
Conservatively taking this additional source of uncertainty into account, the overall uncertainty 
on the background is enlarged to 20\%.

This 20\% is 
added in quadrature to the statistical
uncertainty of the sample to which the fake rate is applied; the latter 
uncertainty 
dominates in the highest two \mee\ bins.
The resulting overall 
uncertainty on the cross-section varies between 1.3\% and 7.9\%, depending on \mee.

Two alternative methods to estimate the  dijet and $W$+jets background are considered as cross-checks.
The first of these is similar to the baseline method but uses fake rates derived from loosely selected electrons 
collected by the EM signal trigger. 
Here the background-enriched sample is derived by employing a tag-and-probe
technique selecting, among other requirements to suppress real electron contamination, a jet-like tag and a probe with the same charge.
This method, being correlated to the baseline method due to the overlap of 
electron candidates passing the EM and jet triggers, yields very similar predictions with comparable systematic uncertainties.
In the  third method, 
the combined dijet plus $W$+jets background is estimated
by performing a template fit to the 
isolation of the leading versus sub-leading electron.
The background templates are obtained from data by 
reversing some of the identification requirements
on one or both of the electrons, and the signal templates
are made from the {\tt PYTHIA} DY sample.
No additional systematic uncertainty is
assigned from the two cross-checks, as
their results are in agreement with the baseline method.

The uncertainties on the diboson and $t\bar{t}$ background expectations include the 
theoretical uncertainties on their cross-sections, 5\% for the
dibosons~\cite{bib:Moc08} and 10\% for 
$t\bar{t}$~\cite{bib:Ali11}. At high~\mee, the statistical uncertainties on the simulated samples dominate,
exceeding 50\% in the highest bin for both processes.
The resulting uncertainty on the cross-section is small compared to the data-driven  dijet and $W$+jets contributions,
ranging from less than 0.3\% at low~\mee~to 2.0\% in the highest~\mee~bin. The 
uncertainty on the cross-section from the total background expectation is between 1.3\% and 8.2\%.

\begin{table}[t]
\setlength{\tabcolsep}{1pt}
\footnotesize
\begin{center}
\begin{tabular}{lcc} 
\hline
\hline
   & \multicolumn{2}{c}{Uncertainty [\%] in \mee\ bin} \\
Source of uncertainty & 116--130\,GeV &
1000--1500\,GeV \\ 
\hline
Total background estimate (Stat.)    & $0.1$ & $7.6$  \\
Total background estimate (Syst.)    & $1.3$ & $3.1$  \\
Electron energy scale \& resolution      & $2.1 $ & $ 3.3 $  \\
Electron identification   &  2.3 & 2.5  \\
Electron reconstruction   &  1.6 & 1.7  \\
Bin-by-bin correction     & 1.5 & 1.5  \\
Trigger efficiency        & $0.8$ &  $ 0.8$ \\
MC statistics ($C_{\rm{DY}}$ stat.)       & $0.7$ & $ 0.4$  \\
MC modelling           & $0.2$ & $0.3$  \\
Theoretical uncertainty  & 0.3 & 0.4   \\ 
\hline
Total systematic uncertainty    & $4.2$ & $ 9.8$   \\
\hline
Luminosity uncertainty & 1.8& 1.8  \\
\hline
Data statistical uncertainty & 1.1 & 50 \\
\hline
\hline
\end{tabular}
\end{center}
\caption{Summary of 
systematic uncertainties on the cross-section measurement, shown for
the lowest and highest bin in \mee. For some sources the lowest or highest 
uncertainty may lie in an intermediate bin. The data statistical uncertainties are also given for comparison.}
\label{tab:systSummary}
\end{table}

\paragraph{Electron reconstruction and identification}

The reconstruction and identification efficiencies of electrons have
been determined previously from data for electrons with $E_{\rm T}$ up to 
50~GeV,  using tag-and-probe methods in vector-boson decays, 
following the prescription of Ref.~\cite{bib:isem}.
To extend the measurement range of the identification efficiency 
in $E_{\rm T}$, a dedicated tag-and-probe measurement is
made using $Z\to e^+e^-$ decays.  
It employs the isolation method, developed in Ref.~\cite{bib:isem} for $W \to e \nu$ 
final states, to estimate the
background contamination.
Here, $\eta$- and $E_{\rm T}$-dependent background template distributions of the isolation
are obtained from data by reversing some of the requirements applied in the electron identification criteria.
The isolation quantity is defined in a similar way to that used in the selection of the leading electron in the signal sample.
The background isolation templates are then normalised to data in the tail of the distributions where no contribution
from signal is expected, both before
and after applying the identification requirements, in order to estimate the background fraction in the probe sample.
The identification efficiencies are found to be consistent with those
obtained by the method of
Ref.~\cite{bib:isem} in the common measurement range, and are stable
for electrons with $E_{\rm T}$ up to 500~GeV. 

The differences between the measured reconstruction and identification
efficiencies and their values in MC simulation are taken 
as $\eta$- and $E_{\rm T}$-dependent scale factors with which the MC-derived $C_{\rm {DY}}$
is corrected. An additional scale factor for the isolation
requirement on the leading electron is also applied.
Varying the scale factors for the electron reconstruction (identification) 
within their systematic uncertainties results in a change in the cross-section  of up to 1.7\% (2.6\%).

\paragraph{Energy scale and resolution}

Both the scale and resolution corrections, estimated from
$Z\to e^+e^-$ events, are varied in the simulation
within their uncertainties.
The overall effect on the cross-section is between 1.0\% and 3.3\%.

\paragraph{Bin-by-bin correction}

The results obtained from the bin-by-bin correction are cross-checked using an 
iterative Bayesian approach~\cite{bayes} and found to be consistent. In addition, a consistency
 test is performed by correcting the
{\tt MC@NLO} signal sample using the {\tt PYTHIA}-derived
$C_{\rm{DY}}$ factor. 
The discrepancy between the sample corrected in this way and the 
true {\tt MC@NLO} sample is about 1.5\%.  This is due to the slightly
different shapes of the \mee\
distribution from the two generators, considered to represent the possible shape 
difference between data and the {\tt PYTHIA} simulation.
This is conservatively added as
a systematic uncertainty on the cross-section in all \mee\ bins.

\paragraph{Trigger efficiency}

Scale factors to account for the difference 
in the  EM signal-trigger efficiency
between data and simulation 
are obtained
by measuring the efficiency in data and MC events using a tag-and-probe method. 
The $Z \to e^+e^-$ events are tagged by selecting events passing a single-electron trigger, thus providing one electron probe
free of trigger bias to test against the signal-trigger requirements.
The scale factors are very close to unity, and the effect on the cross-section of 
varying them within
their systematic uncertainties is approximately 1\%.

\paragraph{MC statistics and MC modelling}

The finite number of events in the MC samples from which the $C_{\rm {DY}}$ factor is derived contribute
an uncertainty of up to 2.4\% on $C_{\rm {DY}}$ and the computed cross-section.
Systematic uncertainties are associated with the use of
the $K$-factors
 and with the reweighting of the {\tt PYTHIA} signal MC events
 in order to better match the 
transverse momentum distribution of the $Z$ bosons and 
the mean number of interactions per bunch crossing in the data. 
The effect of a further reweighting of
 the vertex position distribution in the $z$ direction, not applied by default when calculating  $C_{\rm {DY}}$,
 is also taken as an uncertainty. 
These uncertainties enter into the calculation of $C_{\rm {DY}}$ and result in 
an overall uncertainty on the cross-section of less than 1\%.
Excellent agreement
in the FSR predictions between {\tt PHOTOS} and {\tt SANC}~\cite{sanc,sanc1} has been shown~\cite{sancphotos}
and uncertainties related to the modelling of the detector response to low-energy 
photons from FSR are negligible.

\begin{table}[t!]
\newcommand\T{\rule{0pt}{2.6ex}}
\newcommand\B{\rule[-1.6ex]{0pt}{0pt}}
\setlength{\tabcolsep}{1pt}
\footnotesize
\begin{center}
  \begin{tabular}{ccccc}
\hline 
\hline
\mee~[GeV]  \T  \B &  ${\frac {\rm{d}\sigma}{\rm{d}m_{ee}}}{\rm{(Born)}}$  &  ~${\frac {\rm{d}\sigma}{\rm{d}m_{ee}}}{\rm{(dressed)}}$  &  ~Stat.\,err.\,[\%] & ~Syst.\,err.\,[\%] \\
\hline 
\T 116--130   &   2.24 ${\times 10^{-1}}$                                 &  2.15 ${\times 10^{-1}}$                                   &  1.1        &  4.2\\
\T 130--150   &   1.02 ${\times 10^{-1}}$                                 &  9.84 ${\times 10^{-2}}$                                   &  1.4        &  4.3\\
\T 150--170   &   5.12 ${\times 10^{-2}}$                                 &  4.93 ${\times 10^{-2}}$                                   &  2.0        &  4.6\\
\T 170--190   &   2.84 ${\times 10^{-2}}$                                 &  2.76 ${\times 10^{-2}}$                                   &  2.7        &  4.7\\
\T 190--210   &   1.87 ${\times 10^{-2}}$                                 &  1.82 ${\times 10^{-2}}$                                   &  3.0        &  5.3\\
\T 210--230   &   1.07 ${\times 10^{-2}}$                                 &  1.04 ${\times 10^{-2}}$                                   &  4.4        &  6.1 \\
\T 230--250   &   8.23 ${\times 10^{-3}}$                                 &  7.98 ${\times 10^{-3}}$                                   &  5.2        &  5.9 \\
\T 250--300   &   4.66 ${\times 10^{-3}}$                                 &  4.52 ${\times 10^{-3}}$                                   &  4.3        &  5.8\\
\T 300--400   &   1.70 ${\times 10^{-3}}$                                 &  1.65 ${\times 10^{-3}}$                                   &  5.1        &  5.9 \\
\T 400--500   &   4.74 ${\times 10^{-4}}$                                 &  4.58 ${\times 10^{-4}}$                                   &  9.4        &  6.3\\
\T 500--700   &   1.46 ${\times 10^{-4}}$                                 &  1.41 ${\times 10^{-4}}$                                   &  11         &  5.7\\
\T 700--1000  &   2.21 ${\times 10^{-5}}$                                 &  2.13 ${\times 10^{-5}}$                                   &  24         &  7.5\\
\T 1000--1500 &   2.88 ${\times 10^{-6}}$                                 &  2.76 ${\times 10^{-6}}$                                   &  50         &  9.8\\
\hline                                 
\hline
\end{tabular}
\caption{Measured differential cross-sections  $\frac {{\rm{d}}\sigma}{{\rm{d}}m_{ee}}$ (in pb/GeV) at the Born and dressed levels for DY production of  $e^+ e^-$  pairs in the fiducial region 
(electron $\pt>25$~GeV and $|\eta|<2.5$) with statistical (stat.) and systematic (syst.)
uncertainties in \%. The 1.8\% luminosity uncertainty is not included. }
\end{center}
\label{tab:fid}
\end{table}

\paragraph{Theoretical uncertainties}

Several theoretical uncertainties apply 
to the extrapolation of the cross-section in $|\eta|$ from the measured
region to the fiducial region and thus contribute to an additional uncertainty on
$C_{\rm{DY}}$.
To evaluate the effect of the choice of PDF,
the calculation of $C_{\rm{DY}}$ using {\tt PYTHIA}
with its default PDF ({\tt MRSTMCal}) is compared to that obtained after reweighting to {\tt CT10} (NLO) and 
{\tt HERAPDF1.5}~\cite{herapdf10} (NLO). 
The largest difference between the reweighted results and the
default  is taken as the systematic uncertainty, and amounts to 0.2\%.
A further
systematic uncertainty is calculated using the 
{\tt MC@NLO} sample reweighted to the 52 {\tt CT10} eigenvector error sets, 
the result being 0.5\% at most.
Finally,
comparisons are made between 
{\tt PYTHIA}
reweighted to the {\tt CT10} PDF and {\tt MC@NLO} (which uses as default {\tt CT10}), and
cross-checked using {\tt FEWZ 2.1} at NLO using the {\tt CT10} PDF. 
The effect is at most 0.3\%.
These systematic uncertainties, which each have a
different dependence on \mee, are added in quadrature and together give a 0.2--0.5\% 
uncertainty on the cross-section.

\vskip 2mm

The contributions from the above sources of systematic uncertainty to
the uncertainty on the measured cross-section 
are summarised in Table~\ref{tab:systSummary}
for the lowest and highest bin in the \mee~range considered. 
The overall systematic uncertainty, excluding the luminosity uncertainty of 1.8\%~\cite{lumi}, 
rises from 4.2\% in the lowest \mee~bin to 9.8\% in the highest \mee~bin. 
The data statistical uncertainties increase from 1.1\% to 50\%.

\section{Results and comparison to theory}

\begin{figure}[t]
\begin{center}
\includegraphics[width=0.495\textwidth]{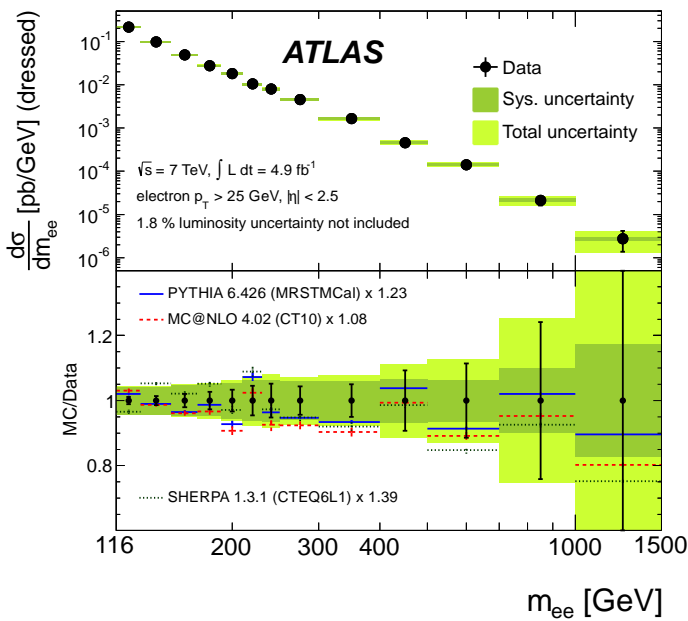}
\end{center}
\caption{Measured differential cross-section at the dressed  level within the 
fiducial region (electron $\pt>25$~GeV and $|\eta|<2.5$) with
statistical, systematic,
and combined statistical and systematic (total) uncertainties, excluding the 1.8\% uncertainty on the
luminosity.
In the lower panel, the measurement is  compared to the predictions 
of the {\tt PYTHIA}, {\tt MC@NLO} and {\tt SHERPA} MC generators including their statistical uncertainties.
No corrections have been applied to the cross-section predictions of the generators. Instead, 
the predictions of each generator have been scaled by a global factor
as indicated on the ratio plots to match the total number of events observed in data.
}
\label{fig:CS_gen}
\end{figure}

The cross-sections obtained
in the fiducial region  (electron $p_{\rm T} > 25$~GeV and $|\eta|<2.5$)
at the Born and dressed levels
are given in Table~2. The difference between the two results is at most 4\%.
The precision of the measurement is limited by the statistical uncertainty
on the data for $\mee > 400$~GeV.

Fig.~\ref{fig:CS_gen} shows the results at the dressed level, where they are compared to the predictions 
from {\tt PYTHIA}, {\tt MC@NLO} and {\tt SHERPA}.
No corrections have been applied
to the generator-level
predictions; instead, the prediction of each generator has been scaled globally to match the total number of events observed in data. 
The resulting scale factors are 1.23 for {\tt PYTHIA}, 1.08 for {\tt MC@NLO} and 1.39 for {\tt SHERPA}.
As expected, the only prediction at NLO in pQCD, from the {\tt MC@NLO} generator,
yields the scale factor closest to unity. 
The overall shape of the \mee\ distribution from all three generators is 
consistent with the data.

\begin{figure*}[t]
\begin{center}
\includegraphics[width=0.495\textwidth]{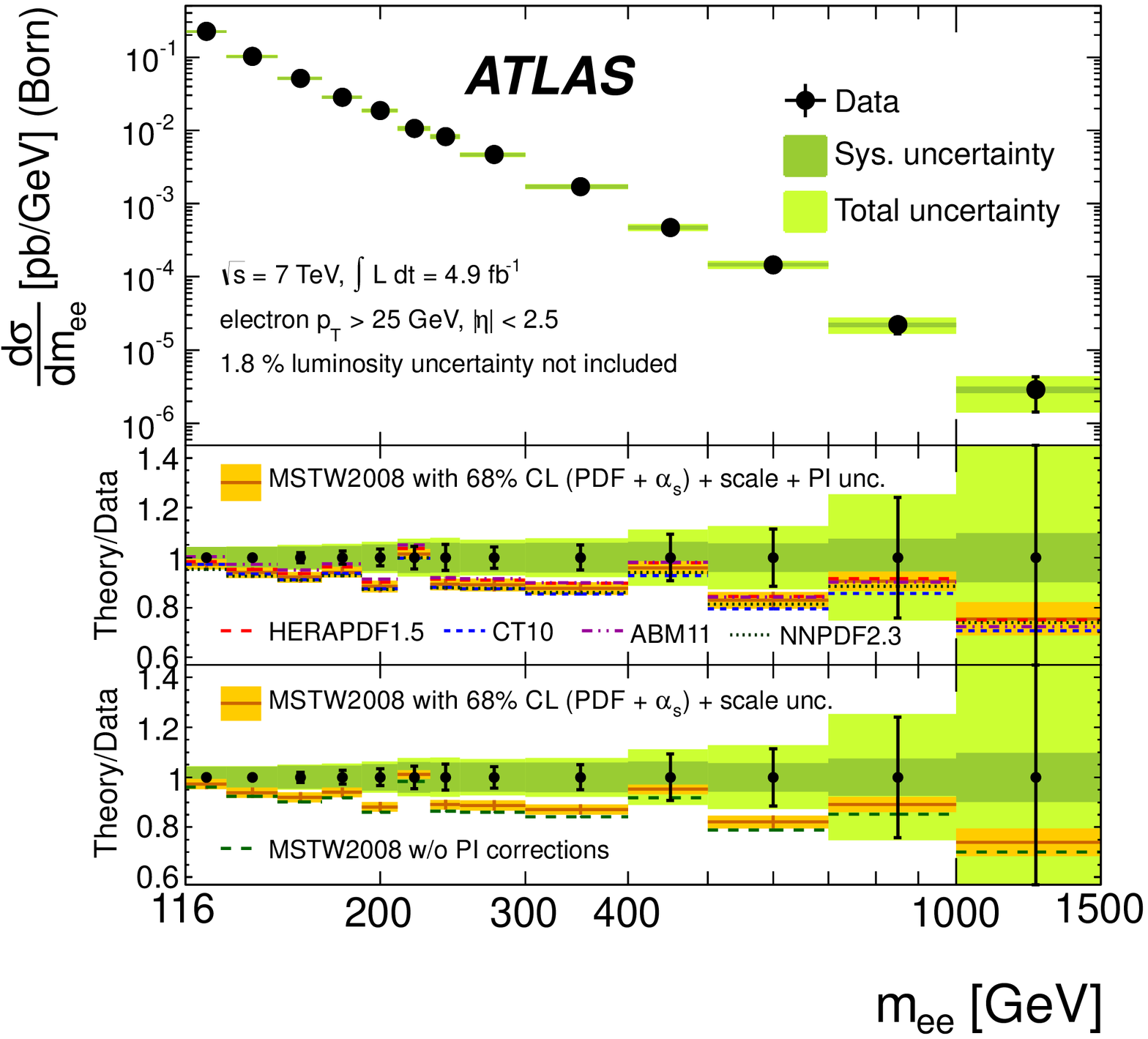}
\includegraphics[width=0.495\textwidth]{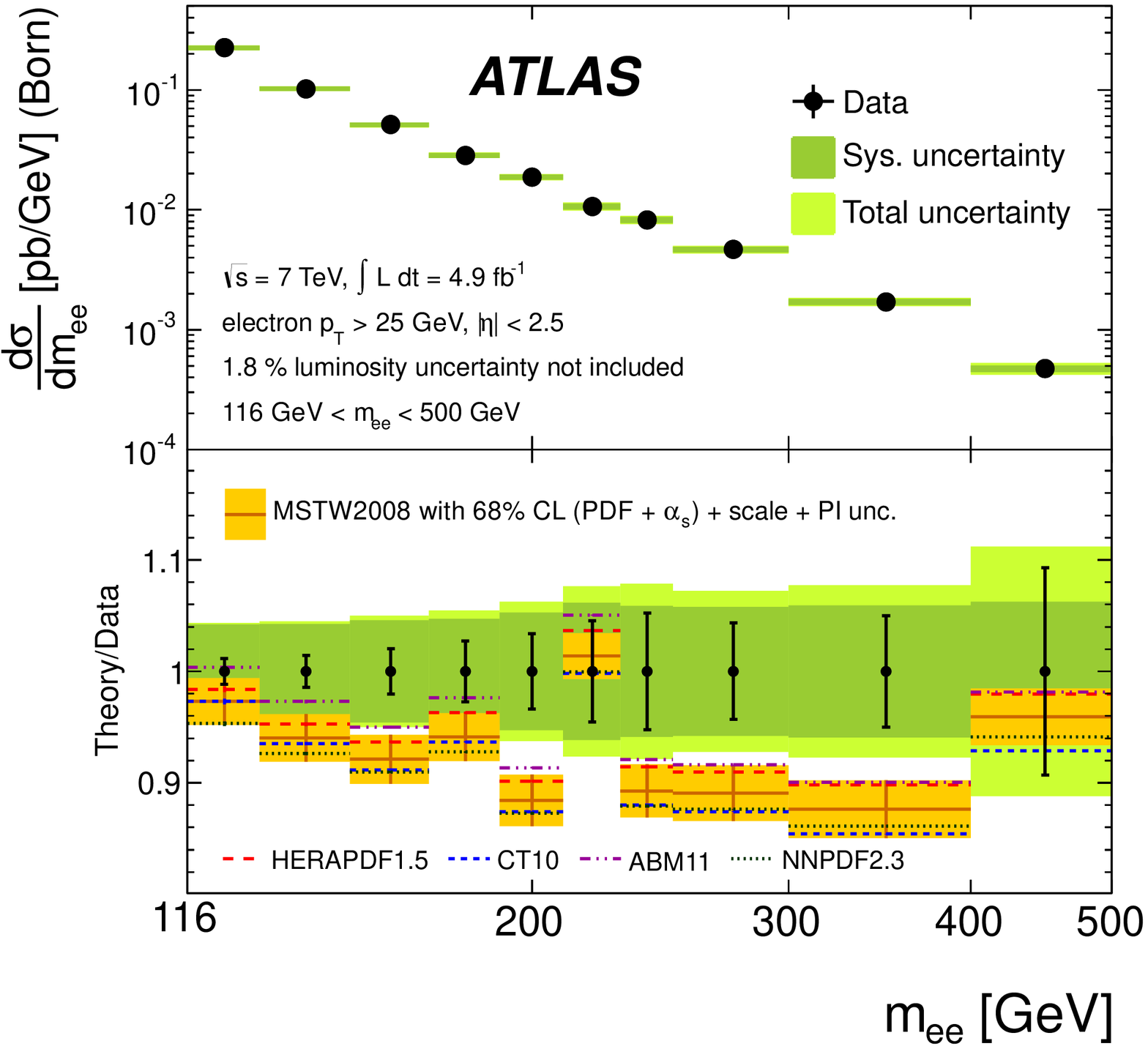}
\end{center}
\caption{Measured differential cross-section at the Born level within the 
fiducial region (electron $\pt>25$~GeV and $|\eta|<2.5$) with
statistical, systematic,
and combined statistical and systematic (total) uncertainties, excluding the 1.8\% uncertainty on the
luminosity.
The measurement is  compared to {\tt FEWZ 3.1} calculations at NNLO QCD with 
NLO electroweak corrections using
the $G_{\mu}$ electroweak parameter scheme. The predictions include an additional small correction from 
single-boson production
in which the final-state charged lepton radiates a real $W$ or $Z$ boson.
On the left, in the upper ratio plot, the photon-induced (PI) corrections
have been added to the  predictions obtained from the {\tt MSTW2008},  {\tt HERAPDF1.5}, {\tt CT10}, {\tt ABM11} and {\tt NNPDF2.3}
NNLO PDFs, and for the {\tt MSTW2008} prediction the total uncertainty band arising from the PDF,
 $\alpha_s$, renormalisation and factorisation scale, and photon-induced uncertainties is drawn. The lower ratio plot
shows the influence of the photon-induced corrections on the {\tt MSTW2008} prediction, the uncertainty
band including only the PDF, $\alpha_s$ and scale uncertainties. On the right, the results are shown 
for a restricted range of \mee.}
\label{fig:CS_calc}
\end{figure*}

Fig.~\ref{fig:CS_calc} shows the differential cross-section at the Born level 
compared to calculations in the 
{\tt FEWZ 3.1} framework using various recent NNLO PDFs.
The {\tt FEWZ 3.1} framework allows the (N)NLO QCD corrections to lepton pair production to be combined with the NLO electroweak
corrections.
It has been verified at NLO in QCD that the choice of electroweak scheme, $G_{\mu}$ 
or $\alpha(m_Z)$ as introduced in Ref.~\cite{Dittmaier:2009cr}, has an
effect of at most 0.4\% on the calculated cross-section after applying NLO electroweak corrections. 
The electroweak-corrected NNLO QCD predictions shown are calculated
using the $G_{\mu}$ scheme.  
The electroweak corrections include a positive contribution
from the irreducible, non-resonant photon-induced background, i.e., $\gamma\gamma \to e^+e^-$. 
This contribution is estimated at leading order (LO) using the {\tt MRST2004qed}~\cite{Martin:2004dh} PDF, currently the only set available that 
includes QED corrections to the proton PDF,
by taking the average of the predictions obtained under the current and constituent quark mass schemes.
The symmetric difference between the average and either scheme is assigned as the corresponding uncertainty on this additive correction, being approximately 50\% and representing a 3\% uncertainty on the cross-section prediction in the highest \mee\ bin.
The electroweak and photon-induced corrections were verified by {\tt SANC}~\cite{sanc,sanc1}.
An additional small correction arises from single-boson production
in which the final-state charged lepton radiates a real $W$ or $Z$ boson~\cite{baur}.
This is estimated using {\tt MADGRAPH 5}~\cite{madgraph}, following the prescription
outlined in Ref.~\cite{baur}, to be at most 2\%, in the highest \mee\ bin.

It can be seen in Fig.~3 
that the deviations  between the 
{\tt MSTW2008}~\cite{mstw2008} and the {\tt CT10}~\cite{bib:ct10},
{\tt HERAPDF1.5}~\cite{herapdf10} and {\tt NNPDF2.3}~\cite{nnpdf23} predictions are
covered by the total uncertainty band assigned to the {\tt MSTW2008} prediction,
which is dominated by the combined
68\% confidence level (CL) PDF and $\alpha_s$ variation.
At low \mee the  {\tt ABM11}~\cite{Alekhin:2012ig} prediction lies above
this theoretical  uncertainty band,
in part due to the  {\tt ABM11} PDF set using a value of $\alpha_s$ outside of the 68\% CL variation.
The renormalisation and factorisation scale uncertainties contribute
at most 1\% to the theoretical uncertainty band
in the highest \mee\ bin, having been evaluated by varying
both scales up or down together by a factor of two, using {\tt VRAP}~\cite{fewzold}.
The size of the photon-induced contribution is similar to the sum of the PDF, $\alpha_s$ and scale uncertainties as can be seen in
the lower panel of Fig.~\ref{fig:CS_calc} (left), where the nominal
calculation using the {\tt MSTW2008} PDF set is compared to the case where
this contribution is not taken into account.

In the region where the precision of the measurement is limited by
systematic uncertainties, $\mee < 400$~GeV, the data generally lie
above the {\tt FEWZ} calculations.
However,  assuming that all systematic uncertainties, except those of
statistical origin on the 
background and on $C_{\rm{DY}}$ (Table~1), are fully correlated bin-to-bin,
the comparison between data and the different
predictions over the full mass range yields chi-squared values of 13.9
for {\tt MSTW2008}, 
18.9 for {\tt CT10}, 13.5 for {\tt HERAPDF1.5}, 14.7 for {\tt ABM11}
and 14.8 for {\tt NNPDF2.3}, for the 13 data points, 
indicating compatibility between the theory and data.

\section{Summary}

Using $4.9$~fb$^{-1}$ of data from  $pp$ collisions at a centre-of-mass energy 
of  $\sqrt{s}=7$~TeV,
the invariant mass distribution of  $e^+ e^-$  pairs from DY production has been measured 
at ATLAS in the range
$116 < \mee < 1500$~GeV, for electrons with
$p_{\rm T} > 25$~GeV and $|\eta| < 2.5$.
Comparisons have been made to the predictions of the {\tt PYTHIA}, {\tt MC@NLO} and {\tt SHERPA} MC generators,
after scaling them globally to match the total number of events observed in data. 
The MC predictions are consistent 
with the shape of the measured \mee\ distribution.
The predictions of the {\tt FEWZ 3.1} framework using
five PDF sets at NNLO 
have also been studied.
The framework combines calculations at NNLO QCD with NLO electroweak corrections,
to which LO photon-induced corrections and real $W$ and $Z$ boson emission in single-boson 
production have been added.
The resulting predictions for all PDFs are consistent with the measured differential cross-section,
although the data 
are systematically above the theory.
The data have the potential to constrain PDFs, in particular for
antiquarks at large $x$, 
in the context of a PDF fit involving the world data sensitive to the proton structure.

\section{Acknowledgements}

We thank CERN for the very successful operation of the LHC, as well as the
support staff from our institutions without whom ATLAS could not be
operated efficiently.

We acknowledge the support of ANPCyT, Argentina; YerPhI, Armenia; ARC,
Australia; BMWF and FWF, Austria; ANAS, Azerbaijan; SSTC, Belarus; CNPq and FAPESP,
Brazil; NSERC, NRC and CFI, Canada; CERN; CONICYT, Chile; CAS, MOST and NSFC,
China; COLCIENCIAS, Colombia; MSMT CR, MPO CR and VSC CR, Czech Republic;
DNRF, DNSRC and Lundbeck Foundation, Denmark; EPLANET, ERC and NSRF, European Union;
IN2P3-CNRS, CEA-DSM/IRFU, France; GNSF, Georgia; BMBF, DFG, HGF, MPG and AvH
Foundation, Germany; GSRT and NSRF, Greece; ISF, MINERVA, GIF, DIP and Benoziyo Center,
Israel; INFN, Italy; MEXT and JSPS, Japan; CNRST, Morocco; FOM and NWO,
Netherlands; BRF and RCN, Norway; MNiSW, Poland; GRICES and FCT, Portugal; MERYS
(MECTS), Romania; MES of Russia and ROSATOM, Russian Federation; JINR; MSTD,
Serbia; MSSR, Slovakia; ARRS and MIZ\v{S}, Slovenia; DST/NRF, South Africa;
MICINN, Spain; SRC and Wallenberg Foundation, Sweden; SER, SNSF and Cantons of
Bern and Geneva, Switzerland; NSC, Taiwan; TAEK, Turkey; STFC, the Royal
Society and Leverhulme Trust, United Kingdom; DOE and NSF, United States of
America.

The crucial computing support from all WLCG partners is acknowledged
gratefully, in particular from CERN and the ATLAS Tier-1 facilities at
TRIUMF (Canada), NDGF (Denmark, Norway, Sweden), CC-IN2P3 (France),
KIT/GridKA (Germany), INFN-CNAF (Italy), NL-T1 (Netherlands), PIC (Spain),
ASGC (Taiwan), RAL (UK) and BNL (USA) and in the Tier-2 facilities
worldwide.

\bibliographystyle{model1-num-names}
\bibliography{Paper_HighMassDY}

\onecolumn
\clearpage
\begin{flushleft}
{\Large The ATLAS Collaboration}

\bigskip

G.~Aad$^{\rm 48}$,
T.~Abajyan$^{\rm 21}$,
B.~Abbott$^{\rm 112}$,
J.~Abdallah$^{\rm 12}$,
S.~Abdel~Khalek$^{\rm 116}$,
A.A.~Abdelalim$^{\rm 49}$,
O.~Abdinov$^{\rm 11}$,
R.~Aben$^{\rm 106}$,
B.~Abi$^{\rm 113}$,
M.~Abolins$^{\rm 89}$
% ...
\bigskip

% ...
$^{11}$ Institute of Physics, Azerbaijan Academy of Sciences, Baku, Azerbaijan\\
$^{12}$ Institut de F\'isica d'Altes Energies and Departament de F\'isica de la Universitat Aut\`onoma de Barcelona and ICREA, Barcelona, Spain\\
% ...
$^{21}$ Physikalisches Institut, University of Bonn, Bonn, Germany\\
% ...
$^{48}$ Fakult\"{a}t f\"{u}r Mathematik und Physik, Albert-Ludwigs-Universit\"{a}t, Freiburg, Germany\\
$^{49}$ Section de Physique, Universit\'e de Gen\`eve, Geneva, Switzerland\\
% ...
$^{89}$ Department of Physics and Astronomy, Michigan State University, East Lansing MI, United States of America\\
% ...
$^{106}$ Nikhef National Institute for Subatomic Physics and University of Amsterdam, Amsterdam, Netherlands\\
% ...
$^{112}$ Homer L. Dodge Department of Physics and Astronomy, University of Oklahoma, Norman OK, United States of America\\
$^{113}$ Department of Physics, Oklahoma State University, Stillwater OK, United States of America\\
% ...
$^{116}$ LAL, Universit\'e Paris-Sud and CNRS/IN2P3, Orsay, France\\
% ...
\end{flushleft}

\end{document}